\documentclass{article}

\usepackage{arxiv}

\usepackage[utf8]{inputenc} 
\usepackage[T1]{fontenc}    
\usepackage{hyperref}       
\usepackage{url}            
\usepackage{booktabs}       
\usepackage{amsfonts}       
\usepackage{nicefrac}       
\usepackage{microtype}      
\usepackage{lipsum}		
\usepackage{graphicx}
\usepackage{natbib}
\usepackage{doi}

\usepackage{amsmath}

\setcitestyle{citesep={;}}

\title{A Theoretical Analysis of Logistic Regression and Bayesian Classifiers}

\author{
    \href{https://orcid.org/0000-0002-9311-8002}{
    \includegraphics[scale=0.06]{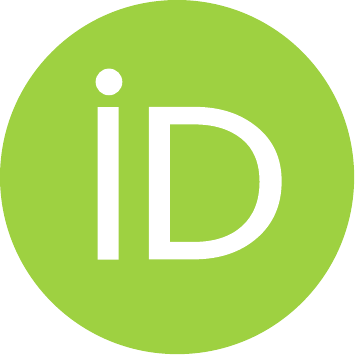}\hspace{1mm}Roman V.~Kirin} \\
	Department of Applied Economics\\
	Higher School of Economics\\
	Pokrovsky blvd. 11, Moscow \\
	\texttt{rkirin@hse.ru}
}

\date{}

\hypersetup{
pdftitle={A Theoretical Analysis of Logistic Regression and Bayesian Classifiers},
pdfsubject={econ, stat},
pdfauthor={Roman V.~Kirin},
pdfkeywords={Bayes’ theorem, distribution families, logistic regression, method of moments},
}

\begin{document}
\maketitle

\begin{abstract}
This study aims to show the fundamental difference between logistic regression and Bayesian classifiers in the case of exponential and unexponential families of distributions, yielding the following findings. First, the logistic regression is a less general representation of a Bayesian classifier. Second, one should suppose distributions of classes for the correct specification of logistic regression equations. Third, in specific cases, there is no difference between predicted probabilities from correctly specified generative Bayesian classifier and discriminative logistic regression.
\end{abstract}


\textbf{\emph{JEL Classification:}} C13, C18, C25, C35.

\textbf{\emph{Keywords:}} Bayes’ theorem, distribution families, logistic regression, method of moments.

\section{Introduction}
The comparison of generative and discriminative models has been a perennial topic of discussion. An example is a comparison of naive Bayes classifier and logistic regression in terms of a classification quality or an error 
\citep{ng2002, gladence2015, ilia2016, prabhat2017, pranckevicius2017, opeyemi2018, setyawan2018, hasanli2019, seka2019, itoo2020}. However, what has been unreasonably ignored in this debate is the fundamental relationship between the models, when both classes have a multivariate normal distribution with similar covariance matrices \citep{efron1975, hastie1997}.

\citet{efron1975} stated that the conditional likelihood for logistic regression ‘… is valid under general exponential family assumptions …’ However, this study shows that it is correct only under the linearity by the features assumption of the discriminant function.

The novelty of this study is that it offers:
\begin{itemize}
\item a generalization for the multiclass task;
\item a solution without a likelihood maximization;
\item a covariance dissimilarity for classes with multivariate normal distribution;
\item an analysis in terms of families of distributions, not prediction accuracy.
\end{itemize}

\section{Problem statement}
\subsection{General case}
Suppose we have a sample $\left\{y_{i\ },\ x_i\right\}_{i=1}^N$ where $y_i\in\left\{1,\ 2,\ ...,\ M\right\}$ denotes the correct class label, and $x_i$ denotes the feature vector of size $d$ for observation $i$. There are two well-known approaches to estimate parameters with maximum likelihood: generative (informative) and discriminative \citep{hastie1997}. Instead of likelihood maximization, let us directly draw the conditional probability for an observation.

Suppose $m$ is a target class, whose conditional probability $P\left(y_i=m\mid x_i\right)$ we want to estimate. According to Bayes’ theorem, the estimated probability can be written as a classifier (\ref{eq1}):

\begin{multline}\label{eq1}
P\left(y_{i}=m\middle|x_{i}\right) 
=\\
\frac{
    P\left(x_{i}\middle|y_{i}=m\right) \times P(y_{i}=m)
}{
    P\left(x_{i}\middle|y_{i}=m\right) \times P\left(y_{i}=m\right)
    +
    \sum_{s\neq m}P\left(x_{i}\middle|y_{i}=s\right) \times P(y_{i}=s)
}=\\
\frac{
    1
}{
    1
    +
    \sum_{s\neq m}
    \frac{
        P\left(x_{i}\middle|y_{i}=s\right) \times P\left(y_{i}=s\right)
    }{
    P\left(x_{i}\middle|y_{i}=m\right) \times P\left(y_{i}=m\right)
    }
}
\end{multline}

The unconditional probability for any class $s$ and observation $i$ equals $P\left(y_i=s\right)=\frac{n_s}{N}$, where $n_s$ is the number of objects with class $s$. The conditional probability $P\left(x_{i}\middle|y_{i}=s\right)$ depends on a specific distribution and a family of the distributions.
\subsection{Exponential families}
Consider that the conditional probability for any class $s$ and any observation $i$ belongs to the exponential family of distributions. Hence, $P\left(x_{i}\middle|y_{i}=s\right)$ can be represented as $P(x_i;\theta_s)=\exp\left(\left\langle t_s\left(x_i\right),\theta_s\right\rangle- F_s\left(\theta_s\right)+k_s\left(x_i\right)\right)$, where $t_s(x_i)$ is the sufficient statistic, $\theta_s$ denotes the natural parameters, $\langle ., . \rangle$ is the inner product (commonly called dot product), $F\left( \cdot \right)$ is the log-normalizer, and $k_s\left(x_i\right)$ is the carrier measure \citep{nielsen2011}.

Under this consideration equation (\ref{eq2}) is appropriate, where log-odds ratio equals (\ref{eq3}) and the log-fraction of unconditional probabilities equals (\ref{eq4}).
\begin{equation}\label{eq2}
P\left(y_{i}=m\middle|x_{i}\right)=\frac{
    1
}{
    1
    +
    \sum_{s\neq m}
    e^{
        -(
            \ln{\rho_{m,s}^{cond}(x_i)}
            +
            \ln{\rho_{m,s}^{uncond}}
        )
    }
}
\end{equation}

\begin{multline}\label{eq3}
\ln{ \rho_{m,s}^{cond} \left(x_i\right) }
=
\ln{
    \frac{
    P\left(x_{i}\middle|y_{i}=m\right)
    }{
    P\left(x_{i}\middle|y_{i}=s\right)
    }
}
=
\left(
\langle t_m\left(x_i\right), \theta_m \rangle
-
\langle t_s\left(x_i\right), \theta_s \rangle
\right)
- 
\left(
F_m\left(\theta_m\right)
-
F_s\left(\theta_s\right) 
\right)
+ 
\left(
k_m\left(x_i\right)
-
k_s\left(x_i\right)
\right)
\end{multline}

\begin{equation}\label{eq4}
\ln\rho_{m,s}^{uncond}=
\ln\frac{P\left(y_i=m\right)}{P\left(y_i=s\right)}=\ln\frac{n_m}{n_s}
\end{equation}
As a result, any Bayesian classifier can be represented as a logistic regression, which was known \citep{efron1975}. But there is no guarantee that the discriminant function $z_{m,s}\left(x_i\right)=\ln\rho_{m,s}^{cond}\left(x_i\right)+\ln\rho_{m,s}^{uncond}$ can be represented as $\beta_0+\beta^Tx_i$.
\section{Particular solutions}
\subsection{Examples for exponential families}
\subsubsection{Univariate normal distribution}
According to the given specification, there are $M$ normally distributed classes, with only one feature ($x_i$ is 1-dimensional). Then, for each class $s$ a number of equations (\ref{eq5}) are valid \citep{nielsen2011}.

\begin{align}\label{eq5}
\begin{split}
\theta_s &= 
\left( \theta_{s, 1} , \theta_{s, 2} \right)
=
\left( \frac{\mu_s}{\sigma^2_s}, -\frac{1}{2\sigma^2_s}  \right)
\\
t_s\left(x_i\right) &= \left( x_i, x_i^2 \right)
\\
F_s\left(\theta_s\right) &= -\frac{\theta_{s, 1}^2}{4\theta_{s, 2}} + \frac{1}{2}\ln{ \left( -\frac{\pi}{\theta_{s, 2}} \right)} 
=
\\
&= 
\frac{1}{2}
\left(
\frac{\mu^2_s}{\sigma^2_s}
\right)
+ \frac{1}{2}\ln{\left(2\pi\right)} + \frac{1}{2} \ln{ \left( \sigma^2_s \right) }
\\
k_s\left(x_i\right) &= 0
\end{split}
\end{align}

The discriminant function for a couple  and  will take this form (\ref{eq6}).

\begin{align}\label{eq6}
\begin{split}
z_{m, s}(x_i) &= \alpha_{m, s} + \beta_{m, s} \times x_i + \gamma_{m, s} \times x_i^2
\\
\alpha_{m, s} &= 
    \ln{\left(\frac{n_m}{n_s}\right)}
    -
    \frac{1}{2}
    \left(
        \ln{\left(\frac{\sigma_m^2}{\sigma_s^2}\right)}
        +
        \left( 
            \frac{\mu_m^2}{\sigma_m^2} - \frac{\mu_s^2}{\sigma_s^2} 
        \right)
    \right) 
\\
\beta_{m, s} &= 
    \left( \frac{\mu_m}{\sigma_m^2} - \frac{\mu_s}{\sigma_s^2} \right)
\\
\gamma_{m, s} &= 
    -
    \frac{1}{2} 
    \left(
        \frac{1}{\sigma_m^2} - \frac{1}{\sigma_s^2}
    \right)
\end{split}
\end{align}
The solution for the specific case leads to the following findings. First, the discriminant function is not linear by feature, in general. Second, if feature variances are the same for both classes, then the logistic regression equations are linear. 

\subsubsection{Multivariate normal distribution}
This section presents the solution for a set of multivariate normally distributed classes. It describes the generalization of the previous subsection when $x_i$ is $d$-dimensional. According to the current task, equations (\ref{eq7}) are valid.

\begin{align}\label{eq7}
\begin{split}
\mathbf{\Theta_s} &=
\left( 
    \theta_s, \Theta_s
\right)
=
\left( 
\Sigma^{-1}_s\mu_s, \frac{1}{2}\Sigma^{-1}_s
\right)
\\
t_s\left(x_i\right) &= \left( x_i, -x_i x_i^{T} \right)
\\
F_s\left(\mathbf{\Theta_s}\right) 
&= 
    \frac{1}{4} tr \left( \Theta_s^{-1} \theta_s \theta_s^{T} \right) 
    - 
    \frac{1}{2} \ln{\left(\det{\Theta_s}\right)} 
    +
    \frac{d}{2} \ln{\left(\pi\right)}
\\
&=
  \frac{1}{2} \mu_s^{T} \Sigma_s^{-1} \mu_s 
  + 
  \frac{1}{2} \ln{ \left( \det{\Sigma_s} \right) }
  +
  \frac{d}{2} \ln{\left(2\pi\right)}
\\
k_s\left(x_i\right) &= 0
\end{split}
\end{align}

For simplicity, let $\Omega_s=\Sigma_s^{-1}$ for any class $s$ and $\omega_{s,j,h}$ be the corresponding element of the matrix $\Omega_s$. Then the discriminant function can be represented as (\ref{eq8}).

\begin{align}\label{eq8}
\begin{split}
z_{m, s}(x_i) &= \alpha_{m,s}+\sum_{j=1}^{d}\beta_{m,s,j}x_{i,j}+\sum_{j=1}^{d}{\sum_{h=1}^{d}\gamma_{m,s,j,h}}x_{i,j}x_{i,h}
\\
\alpha_{m, s} &= 
    \ln{\left(\frac{n_m}{n_s}\right)}
    -
    \frac{1}{2}
    \left(
        \ln{\left(\frac{\det{\Sigma_m}}{\det{\Sigma_s}}\right)}
        +
        \left( 
            \mu_m^{T} \Sigma_m^{-1} \mu_m - \mu_s^{T} \Sigma_s^{-1} \mu_s
        \right)
    \right) 
\\
\beta_{m,s,j} &= \sum_{h=1}^{d}\left(\omega_{m,j,h}\mu_{m,h}-\omega_{s,j,h}\mu_{s,h}\right)\\
\gamma_{m,s,j,h} &= -\frac{1}{2}\left(\omega_{m,j,h}-\omega_{s,j,h}\right)\\
\end{split}
\end{align}

The analytical solution in this section yields two additional findings. First, if features are not independent (at least for one class), then feature interactions (including squares) should be added to the regression equations. Second, if the features are independent and their variance is the same (for all classes), then the regression equations are linear.

\subsection{Example for unexponential families }
\subsubsection{Univariate uniform distribution}

Consider two classes: $x_i\mid y_i=1\sim U\left[a,c\right]$ and $x_i\mid y_i=2\sim U\left[b,d\right]$, where $a<b<c<d$. Then according to the Bayesian classifier (\ref{eq1}), the estimated probability is that the observation i drawn from the class $m$ equals (\ref{eq9}).
\begin{equation}\label{eq9}
P(y_i=1 \mid x_i)=
\begin{cases}
    0,              & \text{if } x_i < a\\
    1,              & \text{if } a \leq x_i < b \\
    \frac{1}{
        1 + 
        \frac{
            P\left(x_{i}\middle|y_{i}=2\right) \times P\left(y_{i}=2\right)
        }{
        P\left(x_{i}\middle|y_{i}=1\right) \times P\left(y_{i}=1\right)
        }
    },              & \text{if } b \leq x_i \leq c \\
    0,              & \text{if } x_i > c
\end{cases}
\end{equation}
Obviously, the task cannot be solved by logistic regression with a simple linear equation, especially for probability 1. As a result, standard logistic regression does not work on an unexponential family, unlike a Bayesian classifier.

\section{Conclusion}

As models, Bayesian classifiers are more general than simple logistic regression. They are appropriate for both exponential and unexponential distribution families (in the cases considered). Under Bayesian classifiers, there always exists a distribution assumption and there can be an additional restriction on parameters, but they should not be naive by default.
The main assumptions of classical logistic regression are: a) exponential distribution family of classes, and b) linearity by features of the discriminant function. Both assumptions do not always hold, but if the first is appropriate and the regression equations are correctly specified, then logistic regression is a useful representation (especially in econometrics) and leads to a correct solution.
Moreover, learning Bayesian classifier is more computationally efficient because moment’s method can be used. For logistic regression, only variations of gradient ascent are available.

\section*{Conﬂict of Interest}
None.

\section*{Funding}
This research did not receive any speciﬁc grants from funding agencies in the public, commercial, or not-for-profit sectors.

\bibliographystyle{unsrtnat}
\bibliography{references}

\begin{thebibliography}{13}
\providecommand{\natexlab}[1]{#1}
\providecommand{\url}[1]{\texttt{#1}}
\expandafter\ifx\csname urlstyle\endcsname\relax
  \providecommand{\doi}[1]{doi: #1}\else
  \providecommand{\doi}{doi: \begingroup \urlstyle{rm}\Url}\fi

\bibitem[Ng and Jordan(2002)]{ng2002}
Andrew Ng and Michael Jordan.
\newblock {On Discriminative vs. Generative Classifiers: A comparison of
  logistic regression and naive Bayes}.
\newblock \emph{Adv. Neural Inf. Process. Sys}, 2, 04 2002.

\bibitem[L.~Mary~Gladence and Anu(2015)]{gladence2015}
M.~Karthi L.~Mary~Gladence and V.~Maria Anu.
\newblock {A statistical comparison of logistic regression and different bayes
  classification methods for machine learning}.
\newblock \emph{ARPN Journal of Engineering and Applied Sciences}, 10, 2015.

\bibitem[Tsangaratos and Ilia(2016)]{ilia2016}
Paraskevas Tsangaratos and Ioanna Ilia.
\newblock {Comparison of a logistic regression and Naïve Bayes classifier in
  landslide susceptibility assessments: The influence of models complexity and
  training dataset size}.
\newblock \emph{CATENA}, 145:\penalty0 164--179, 2016.
\newblock ISSN 0341-8162.
\newblock \doi{https://doi.org/10.1016/j.catena.2016.06.004}.
\newblock URL
  \url{https://www.sciencedirect.com/science/article/pii/S0341816216302090}.

\bibitem[Prabhat and Khullar(2017)]{prabhat2017}
Anjuman Prabhat and Vikas Khullar.
\newblock {Sentiment classification on big data using Naïve bayes and logistic
  regression}.
\newblock In \emph{2017 International Conference on Computer Communication and
  Informatics (ICCCI)}, pages 1--5, 2017.
\newblock \doi{10.1109/ICCCI.2017.8117734}.

\bibitem[Pranckevicius and Marcinkevičius(2017)]{pranckevicius2017}
Tomas Pranckevicius and Virginijus Marcinkevičius.
\newblock {Comparison of Naive Bayes, Random Forest, Decision Tree, Support
  Vector Machines, and Logistic Regression Classifiers for Text Reviews
  Classification}.
\newblock \emph{Baltic Journal of Modern Computing}, 5, 01 2017.
\newblock \doi{10.22364/bjmc.2017.5.2.05}.

\bibitem[Aborisade and Anwar(2018)]{opeyemi2018}
Opeyemi Aborisade and Mohd Anwar.
\newblock {Classification for Authorship of Tweets by Comparing Logistic
  Regression and Naive Bayes Classifiers}.
\newblock In \emph{2018 IEEE International Conference on Information Reuse and
  Integration (IRI)}, pages 269--276, 2018.
\newblock \doi{10.1109/IRI.2018.00049}.

\bibitem[Helmi~Setyawan et~al.(2018)Helmi~Setyawan, Awangga, and
  Efendi]{setyawan2018}
Muhammad~Yusril Helmi~Setyawan, Rolly~Maulana Awangga, and Safif~Rafi Efendi.
\newblock {Comparison Of Multinomial Naive Bayes Algorithm And Logistic
  Regression For Intent Classification In Chatbot}.
\newblock In \emph{2018 International Conference on Applied Engineering
  (ICAE)}, pages 1--5, 2018.
\newblock \doi{10.1109/INCAE.2018.8579372}.

\bibitem[Hasanli and Rustamov(2019)]{hasanli2019}
Huseyn Hasanli and Samir Rustamov.
\newblock {Sentiment Analysis of Azerbaijani twits Using Logistic Regression,
  Naive Bayes and SVM}.
\newblock In \emph{2019 IEEE 13th International Conference on Application of
  Information and Communication Technologies (AICT)}, pages 1--7, 2019.
\newblock \doi{10.1109/AICT47866.2019.8981793}.

\bibitem[Seka et~al.(2019)Seka, Bonny, Yoboué, Sié, and
  Adopo-Gourène]{seka2019}
D.~Seka, B.S. Bonny, A.N. Yoboué, S.R. Sié, and B.A. Adopo-Gourène.
\newblock {Identification of maize (Zea mays L.) progeny genotypes based on two
  probabilistic approaches: Logistic regression and naïve Bayes}.
\newblock \emph{Artificial Intelligence in Agriculture}, 1:\penalty0 9--13,
  2019.
\newblock ISSN 2589-7217.
\newblock \doi{https://doi.org/10.1016/j.aiia.2019.03.001}.
\newblock URL
  \url{https://www.sciencedirect.com/science/article/pii/S2589721719300030}.

\bibitem[Itoo et~al.(2020)Itoo, Meenakshi, and Singh]{itoo2020}
Fayaz Itoo, Meenakshi, and Satwinder Singh.
\newblock {Comparison and analysis of logistic regression Naïve Bayes and
  {KNN} machine learning algorithms for credit card fraud detection}.
\newblock \emph{International Journal of Information Technology}, feb 2020.
\newblock \doi{10.1007/s41870-020-00430-y}.
\newblock URL \url{https://doi.org/10.1007%2Fs41870-020-00430-y}.

\bibitem[Efron(1975)]{efron1975}
Bradley Efron.
\newblock {The Efficiency of Logistic Regression Compared to Normal
  Discriminant Analysis}.
\newblock \emph{Journal of the American Statistical Association}, 70\penalty0
  (352):\penalty0 892--898, 1975.
\newblock \doi{10.1080/01621459.1975.10480319}.

\bibitem[Rubinstein and Hastie(1997)]{hastie1997}
Y.~Dan Rubinstein and Trevor Hastie.
\newblock {Discriminative vs Informative Learning}.
\newblock In \emph{Proceedings of the Third International Conference on
  Knowledge Discovery and Data Mining}, KDD'97, page 49–53. AAAI Press, 1997.

\bibitem[Nielsen and Garcia(2011)]{nielsen2011}
Frank Nielsen and Vincent Garcia.
\newblock {Statistical exponential families: A digest with flash cards}, 2011.

\end{thebibliography}

\end{document}